\documentstyle[preprint,prb,aps,eqsecnum]{revtex}
\begin{document}
\draft
\input{psfig}
\psfull
\title{An Upsilon Point in a Spin Model}
\author{C. Micheletti$\null^1$, F. Seno$\null^2$ and J. M. Yeomans$\null^1$}
\address{ (1) Theoretical Physics, Oxford University,
1 Keble Rd., Oxford OX1 3NP, UK}
\address{(2) INFM-Dipartimento di Fisica ``G. Galilei'',
Via Marzolo 8, I-35131 Padova, Italy}
\date{\today}
\maketitle
\begin{abstract}
We present analytic evidence for the occurrence of an upsilon point, an
infinite checkerboard structure of modulated phases, in
the ground state of a spin model. The structure of the
upsilon point is studied
by calculating interface--interface interactions using an expansion in
inverse spin anisotropy.

\end{abstract}

\section{Introduction}

Simple spin models can have surprisingly complex phase structures,
even at zero temperature. In particular, near multiphase lines, along
which the ground state is infinitely degenerate, a perturbation such
as temperature [\onlinecite{FS}], quantum fluctuations
[\onlinecite{HMY}] or softening of the spins [\onlinecite{1/d}] can
result in infinite sequences of stable phases.

The existence of upsilon-points ($\Upsilon$-points) in Frenkel-Kontorova models
has been pointed out recently [\onlinecite{CG},
\onlinecite{SF}]. These occur when two multiphase
lines meet at a first-order boundary. A small perturbation about such
a point can stabilise an infinite checkerboard
structure of commensurate phases as shown in Fig. \ref{fig:upsi}. In
many ways the $\Upsilon$-point can be
thought of as a two-dimensional generalisation of the behaviour
customary near a multiphase point.

The occurrence of an $\Upsilon$-point in a spin model was recently
suggested by the numeric work of Sasaki [\onlinecite{Sasaki}]. Here we
present the first analytic evidence
for the existence of a $\Upsilon$-point in a spin model. The system we
consider is the chiral XY model with 6-fold spin anisotropy
in a magnetic field [\onlinecite{Yokoi}]. We identify a candidate
for a $\Upsilon$-point at infinite spin anisotropy $D$  and
show that, as $D$ is reduced from infinity, the softening of the potential
wells allows formation of a $\Upsilon$-point structure.

We follow the method introduced by Fisher and Szpilka
[\onlinecite{FSzI}]
and extended by Bassler, Sasaki, and Griffiths [\onlinecite{Bass}] and identify
the multiphase structures as comprising interfaces separating domains of
the different phases. The behaviour near a single multiphase point
can be analysed in terms of a unique type of interface.
However  there are two (or
more) different phases stable near an $\Upsilon$-point and
hence different types of interface must be identified in the analysis. It
is the interactions between the interfaces which are responsible for
breaking the multiphase degeneracy and a knowledge of their sign and dependence
on separation allows determination of the phase diagram. Here the
interface-interface interactions are
calculated using an expansion in $D^{-1}$ [\onlinecite{1/d}].

\section{The Model}

We consider the classical chiral XY model with 6-fold spin anisotropy, $D$,
in the presence of an external magnetic field $h$.
The Hamiltonian of the system is
\begin{equation}
{\cal H} = \sum_i \biggl\{ - \cos ( \theta_{i-1}- \theta_{i} +
\pi \Delta/3) + h [1 - \cos (\theta_i)]+ {D} [ 1 - \cos (6
\theta_i)] /36  \biggr\}
\label{eqn:ham}
\end{equation}
\noindent where $\theta_i$ is the angle between the $i^{\rm th}$ spin and the
magnetic field orientation. We shall concentrate on the behaviour of
the  model near the
limit $D=\infty$, where $n_i$, defined as $3 \theta_i /\pi$, can
take only the integer values
\{0,1...,5\}.

The ground-state configurations satisfy

\begin{equation}
{\partial{\cal H} \over \partial \theta_i}=0 \quad \forall i.
\label{eqn:rec}
\end{equation}

\noindent For a given $i$, equation (\ref{eqn:rec}) enables us to express
$\theta_{i+1}$ as a function of $\theta_i$ and $\theta_{i-1}$. This fact,
together with the observation that $n_i \in \{0,1...,5\}$, is
sufficient to conclude that, for
$D=\infty$, there will always exist periodic minimal energy configurations.
It will be convenient to label a periodic configuration $\{ ...,
\theta_N,\theta_1,\theta_2,...\theta_N,\theta_1,...\}$ as $ \langle
n_1\, n_2\,...  n_N \rangle$.

\noindent We can now discuss the phase diagram for $D=\infty$, obtained
using the Floria-Griffiths algorithm  [\onlinecite{FG}], and presented
in Fig. \ref{fig:pd}. We have
restricted the labelling of the phases to the first quadrant
$(0 \le \Delta \le 3; h \ge 0)$; the remaining phases can be constructed
through appropriate symmetry operations on the $n_i$ sequences.
\noindent The transition lines between regions A and J and
regions J and F are first order.
The remaining boundaries are multiphase lines, that is loci where all phases
(including non-periodic ones) built from arbitrary combinations of the
two neighbouring phases are degenerate [\onlinecite{FS}].

If the spin anisotropy is reduced from $\infty$ it seems
natural to expect the
degeneracy along the multiphase lines to be lifted as the spins
soften from the clock positions. Although $n_i$ is no longer constrained
to assume integer values, nevertheless, for high values of $D$, the
angles $\theta_i$ will be close enough to the clock positions to allow
us to continue to use the same labelling scheme.

\begin{figure}
\centerline{\psfig{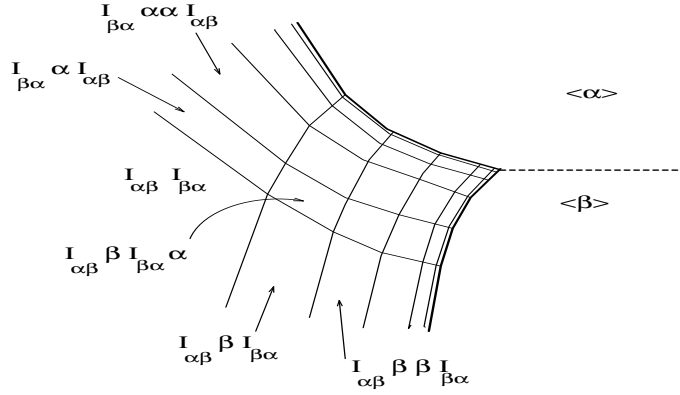}}
\vspace{0.3cm}
\caption{Schematic representation of an $\Upsilon$-point.}
\label{fig:upsi}
\end{figure}

\begin{figure}
\centerline{\psfig{figure=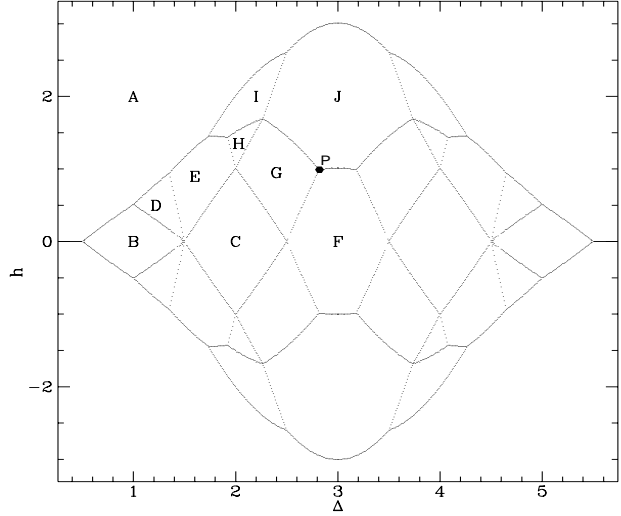,width=3.5in,height=3.0in}}

\caption{Ground state of the Hamiltonian (2.1) for $D=\infty$.
 A $= \langle0\rangle$; B $= \langle0 1 2 3 4 5\rangle $;
C $= \{\langle0 2 4\rangle\  , \langle1 5 3\rangle\}$;
D $=\langle0 1 2 4 5\rangle$;  E $= \langle0 1 3 5\rangle$;  F $=
\{ \langle0 3\rangle\ , \langle1 4\rangle\ , \langle2 5\rangle\}$;
G $= \langle0 2 5 1 4\rangle$;  H $= \{ \langle0 2 5\rangle, \langle0
1 4\rangle \} $;
I $= \langle0 1 5\rangle$;  J $= \langle1 5\rangle$}
\label{fig:pd}
\end{figure}

We are particularly interested in the possible appearance of
$\Upsilon$-points for finite $D$. An $\Upsilon$-point can occur when a
first-order transition line
separating, say, phases $\langle \alpha \rangle$ and $\langle \beta
\rangle$ (that, for simplicity, we now assume to be non-degenerate)
approaches a commensurate-incommensurate transition.
An infinite number of phases spring out from the multicritical
point at the end of the first-order line (as represented in Fig.
\ref{fig:upsi}).
The phases appearing are made of sequences of $\langle \alpha \rangle$
and  $\langle \beta \rangle$.
As $\alpha \not= \beta$ the interfaces separating them, which we shall call
$I_{\alpha \beta}$ and $I_{\beta \alpha}$ are also generally distinct.
In the example of Fig. \ref{fig:upsi} the general form for a phase
in the fan is $\langle \alpha^n I_{\alpha \beta} \beta^m I_{\beta
\alpha}\rangle$, where the integers $n,m$ increase approaching the
$\langle \alpha \rangle$
and $\langle \beta \rangle$ boundaries respectively.

The multiphase point {\bf P} highlighted in Fig. \ref{fig:pd} seems to be a
good
candidate for
becoming an $\Upsilon$-point when $D$ is relaxed from $\infty$.
{\bf P} lies at the end of a first-order transition line and
it seems reasonable to consider the two multiphase lines J-G and G-F
as special cases of accumulation lines.
Therefore we might expect to observe a structure similar to Fig.
\ref{fig:upsi} for small
values of $1/D$.

\section{THE $\langle \alpha \rangle$ BOUNDARY}

Consider the J--G boundary (at a finite distance
from the point {\bf P}). When $D=\infty$ the phases  $\langle \alpha
\rangle \equiv \langle 5 1 \rangle$
and $\langle 5 1 4 0 2 \rangle$ coexist, and it is easy to check that
along the boundary all phases built with $\alpha$ sequences separated
by a $| 4 0 2 | \equiv I$ block (i.e. $\langle \alpha^n I \alpha^m
... \rangle$) are degenerate. We want to study how this degeneracy is
lifted when $D$ assumes finite values.

It is physically appealing to regard the $I$ block of spins as an
interface separating pure $\alpha$ sequences.
Indeed one can conveniently write the energy per spin of, say, phase
$\langle \alpha^n I \rangle$ as
\begin{equation}
E = E^0_I + ( 2n (E^0_{\alpha} - E^0_{I}) +
\sigma +V_{\alpha} (2n) +  V_{\alpha \alpha}(2n,2n)+ ...)/(n_I+2n)
\label{eqn:energy}
\end{equation}
\noindent where $n_I = 3$, $E^0_{I}$
($E^0_{\alpha}$) is the energy per spin of phase $\langle I \rangle$
($\langle \alpha \rangle$), $\sigma$ is the creation energy of $I$,
$V_\alpha(2n)$ is the interaction energy of two interfaces $I$
separated by a distance $2n$,
$V_{\alpha\alpha}(2n,2n)$ is the interaction energy of three
interfaces and so forth.
In the $\langle \alpha \rangle$ region the interface tension $\sigma$
is positive; as the phase boundary is approached $\sigma$ decreases
and eventually, when it is balances the interface interactions, it will be
favourable for the system to replace the pure $\alpha$ phase with a
modulated one. The nature of the transition depends on the form of the
interface interactions, which we now calculate to leading order using
an expansion in inverse spin anisotropy.

\begin{figure}
\centerline{\psfig{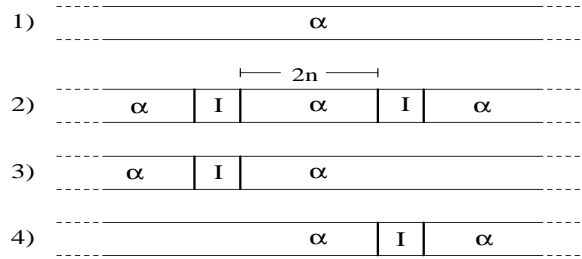}}
\vspace{0.3cm}
\caption{Configurations needed to calculate the two-interface interaction;
see equation (3.2).}
\label{fig:reco}
\end{figure}

In the large $D$ limit $V_{\alpha}(2n)$
dominates the energy contribution from the interface interaction
terms. It can be obtained using the reconnection formula [\onlinecite{Bass}]
\begin{equation}
V_{\alpha}(2n) = E_1 + E_2 - E_3 - E_4
\label{eqn:reco}
\end{equation}
\noindent where $E_i$ is the energy of configuration $i$ sketched in Fig.
\ref{fig:reco}.

\begin{figure}
\centerline{\psfig{figure=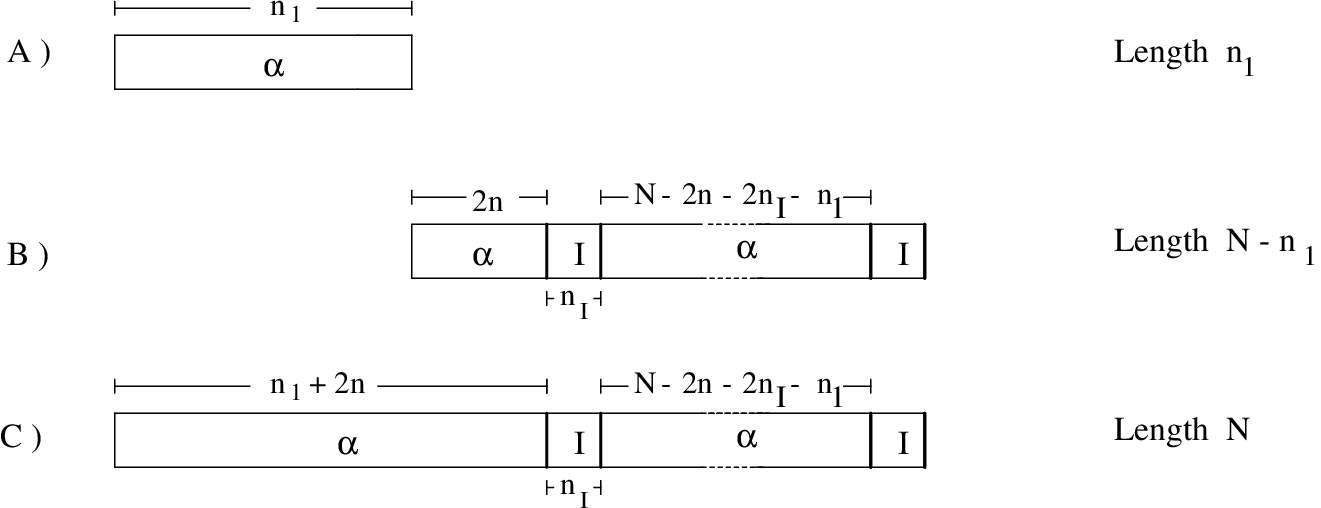,height=2.5in }}
\vspace{0.3cm}
\caption{Periodic configurations appearing in the approximate reconnection
formula (3.3).}
\label{fig:reco2}
\end{figure}

Equation (\ref{eqn:reco}) is exact, but is not convenient for our
purposes, as we want only the leading term of $V_{\alpha}(2n)$.
In fact, we can exploit the rapid decay of the $V_{\alpha}$
with $n$ to substitute all
infinite segments in Fig. \ref{fig:reco} with finite (though
sufficiently long) ones. Thus equation (3.2) can be approximated by
\begin{equation}
V_{\alpha}(2n) \approx E_{\rm A} + E_{\rm B} -
E_{\rm C}
\label{eqn:recon}
\end{equation}
where ${\rm A}_1$,   ${\rm B}$, and  ${\rm C}$, are the {\em
periodic\/} configurations sketched in Fig.\ \ref{fig:reco2}.
$n_1+2n$ and $N-2n-2n_I -n_1$ are assumed to be both a large multiple
of $2n$ and much greater than $n_I$.

We label the spins of configurations A, B and C as
$\{a_1,a_2, ... , a_{n_1} \equiv a_0\}$, $\{b_{n_1+1}, ..., b_N \equiv b_{n_1}
\}$ and $\{c_1, c_2,
... , c_N \equiv c_0\}$ respectively. For $D$ large the spins will deviate from
their clock positions $\{ a^0_i\}$, $\{ b^0_i\}$ and $\{ c^0_i\}$ by an angle
analytic in $D^{-1}$ and we write
\begin{equation}
a_i = a^0_i + \tilde{a}_i, \ \ \ \ \  b_i=b^0_i + \tilde{b}_i, \ \ \ \ \
c_i=c^0_i + \tilde{c}_i  .
\end{equation}

We can choose to label the spins such that
\begin{equation}
a^0_i=c^0_i \ \ \ \ \ \   1 \le i \le n_1, \hskip 2.8cm b^0_i=c^0_i \ \ \ \ \ \
 n_1+1 \le i \le N.
\label{eqn:a}
\end{equation}
\noindent Then using a superscript tilde to indicate we are
working only to second order in the spin deviations $\{\tilde{a}_i\}$,
$\{\tilde{b}_i\}$, $\{\tilde{c}_i\}$ the two-interface interaction can
be written
\begin{eqnarray}
\tilde{V}_\alpha(2n)& =& \tilde{\cal H}(\tilde{a}_{n_1},\tilde{a_1})
+\sum_{i=2}^{n_1} \tilde{{\cal H}} (\tilde{a}_{i-1}, \tilde{a}_i)
+\tilde{\cal H}(\tilde{b}_N,\tilde{b}_{n_1+1}) +
\sum_{i=n_1+2}^{N} \tilde{{\cal H}}(\tilde{b}_{i-1}, \tilde{b}_i)
\nonumber \\
&& \ \ \ \ \ \ \ \ \ \ \ \ \ \ \
-\tilde{\cal H}(\tilde{c}_N,\tilde{c}_1) +
\sum_{i=2}^{N} \tilde{{\cal H}}(\tilde{c}_{i-1}, \tilde{c}_i)
\label{eqn:b}
\end{eqnarray}
\noindent where
\begin{equation}
\tilde{{\cal H}} (\tilde{a}_{i-1}, \tilde{a}_i) = J_{i-1,i}^a
\{ \tilde{a}_{i-1} - \tilde{a}_i + \Delta^a_{i-1,i} \}^2 + h_i^a
(\tilde{a}_i + \epsilon^a_i)^2 + D \tilde{a}_i^2/2
\end{equation}
\noindent with
\begin{eqnarray}
J_{i-1,i}^a &=& \cos (a^0_{i-1} - a^0_i + \pi \Delta /3)/2, \label{eqn:first}\\
h^a_i &=& h \cos (a_i^0)/2, \\
\Delta_{i-1,i}^a &=& \tan ( a^0_{i-1} - a^0_i + \pi \Delta /3), \\
\epsilon_i^a &=& \tan(a^0_i) \label{eqn:last}.
\end{eqnarray}
It follows from (\ref{eqn:a}) that
\begin{equation}
\epsilon_i^a = \epsilon_i^b,\ \ \ \ \  \ h_i^a=h_i^b
\end{equation}
\noindent for all $i$ and that
\begin{eqnarray}
J_{i-1,i}^a &=& J_{i-1,i}^c, \ \ \ \ \
\Delta_{i-1,i}^a=\Delta_{i-1,i}^c,  \ \ \ \ \  \ 2 \le i \le n_1
\label{eqn:JDelta1}\\
J_{i-1,i}^b &=& J_{i-1,i}^c, \ \ \ \ \
\Delta_{i-1,i}^b=\Delta_{i-1,i}^c,  \ \ \ \ \  \ n_1+2 \le i \le N.
\label{eqn:JDelta2}
\end{eqnarray}
For
the cases considered here it  will be possible to label the phases so
that (\ref{eqn:JDelta1}) is also true for $i =1$ and
(\ref{eqn:JDelta2}) for  $i=n_1+1$.
Under these circumstances  we may drop the $a,\ b$ and $c$ superscripts on the
quantities defined in
(\ref{eqn:first})-(\ref{eqn:last}). It is then possible to use the
recursion equations
(\ref{eqn:rec}) to simplify (\ref{eqn:b}). After some algebra one obtains
\begin{equation}
\tilde{V}_\alpha(2n) =- J_{0,1} \{(\tilde{a}_{n_1} - \tilde{b}_N)
(\tilde{c}_1 - \tilde{c}_{n_1+1})-(\tilde{a}_1 -
\tilde{b}_{n_1+1})(\tilde{c}_N - \tilde{c}_{n_1})\}.
\label{eqn:short}
\end{equation}

The quantities appearing in (\ref{eqn:short}) can be obtained to
leading order in $1/D$ using the recursion equation
(\ref{eqn:rec}). An example of how
to calculate $\tilde{V}_{\alpha}(2n)$ is given in Appendix A. The result for
general $n$ is
\begin{equation}
\tilde{V}_{\alpha}(2n) = {\rm c}_2^{n}
{\rm c}_4^{n-1} \bigl\{ {\rm s}_4 -{\rm s}_3 \bigr\}^2 / D^{2n} +
{\cal O}(1/D^{(2n +1)})
\label{eqn:V}
\end{equation}
\noindent where ${\rm s}_i \equiv \sin[\pi  (\Delta
-i)/3]$ and ${\rm c}_i \equiv \cos[\pi (\Delta
-i)/3]$. Terms of higher order than quadratic in the Hamiltonian
(\ref{eqn:ham}) will not contribute to the leading term of the
interface--interface interaction and hence to leading order
$V_{\alpha}(2n)$ and $\tilde{V}_{\alpha}(2n)$ will be equal.
Therefore we shall not distinguish between them below.

A knowledge of the leading term in the interface-interface interaction,
equation (\ref{eqn:V}), allows us to take the first step in determining
the ground-state configurations. Because we are considering only
two-interface interactions the interfaces must be equispaced in the ground
state.
Inspection of equation (\ref{eqn:V}) shows that $V_{\alpha}(2n)$ is always
positive and convex near {\bf P}.
This is enough to
conclude that, for $D$ large,  all
transitions $\langle \alpha^n I \rangle \to \langle \alpha^{n+1} I
\rangle$ occur as $\sigma$ is lowered [\onlinecite{FSzI}].

To this order of approximation the $\langle \alpha^n I \rangle :
\langle \alpha^{n+1} I \rangle$ phase boundaries remain degenerate and
higher-order interface interactions can introduce qualitative changes
in the phase diagram. This will be discussed further in section V.

\section{The $\langle \beta \rangle$ Boundary}

We now focus our attention on what happens along the $\langle \beta
\rangle$ boundary
in the two-interface interaction approximation.
In the F region of the phase diagram (Fig. \ref{fig:pd}) three
phases coexist when $D=\infty$, namely $\beta_1= \langle 1 4 \rangle$,
$\beta_2= \langle 2 5 \rangle$ and $\beta_3=\langle 0 3 \rangle$.
However, when $D$ is relaxed, only phases $\beta_1$ and $\beta_2$
continue to stay degenerate, while phase $\beta_3$ has a higher
energy.

Consider the boundary between one of the phases
$\langle \alpha^{n} I \rangle$ and region F. Along this boundary, in
the absence of interactions between the interfaces $I_1\equiv (51)^n
4,\ I_2 \equiv 0$ and $I_3\equiv 2$, all
phases $\langle (51)^n 4 (14)^{m_1} 0 (30)^{l_1} 2 (52)^{p_1} (51)^n 4
(14)^{m_2}  0 (30)^{l_2} 2 (52)^{p_2} .... \rangle$ are
degenerate.

Now we turn on the two-interface interactions.
In this approximation, the
possible ground states are periodic and have the form $\langle
(51)^n 4 (14)^m 0 (30)^l 2 (52)^p \rangle$, where $m,\ l$ and $p$
depend on $\sigma$.
In the following analysis we shall hold $n$ fixed and assume that
$\sigma$ can be varied to trace out the phase sequences.

The energy per spin can be written

\begin{eqnarray}
E = \biggl\{& & (1+2n) E_{I_1}+E_{\beta_1} 2m + E_{I_2}+
E_{\beta_2} 2p +E_{I_3}+ E_{\beta_3} 2l + \sigma \nonumber \\
& &+V_{\beta_1}(2m) +  V_{\beta_3}(2l)+  V_{\beta_2}(2p)  \biggr\}/L
\label{eqn:ebeta}
\end{eqnarray}

\noindent where $L=(2m+2p+2l+3+2n)$ and $\sigma$ includes the energy tension of
the three interfaces $I_1$, $I_2$ and $I_3$.

\noindent Simple calculations show that
\begin{equation}
E_{\beta_1}=E_{\beta_2},\ \ \
E_{\beta_3}= E_{\beta_2}+ {3 h^2/ (8 D)} + {\cal O}(1/D^2).
\label{eqn:pen}
\end{equation}

\noindent Proceeding as in Section III and Appendix A the two-wall interactions
between interfaces bounding phases $\langle \beta_1 \rangle$, $\langle
\beta_2 \rangle$, and  $\langle
\beta_3 \rangle$ are to leading order

\begin{eqnarray}
 V_{\beta_1}(2m)=
 V_{\beta_2}(2p) &=& D^{-(2m+2)} \bigl\{{\rm s}_2-{\rm
 s}_3\bigr\}^2 {\rm c}_3^{2m+1}, \nonumber \\
 V_{\beta_3}(2l)&=&D V_{\beta_1}(2l)/c_3.
\label{eqn:Vbeta}
\end{eqnarray}

We now want to find the values $\bar{m},\ \bar{p}$ and $\bar{l}$ which
minimise (\ref{eqn:ebeta}) for a given $n$ and $\sigma$.
By symmetry arguments one has $\bar{m}=\bar{p}$. It follows from
(\ref{eqn:pen}) that $\bar{l}$ must be bounded from
above. Indeed an explicit calculation of the energy ${\cal O}(1/D)$
shows immediately that $\bar{l}=0$ or 1 and that the sequence of
phases as $\sigma$ is lowered is, using the notation
$[\bar{n}, \bar{m}, \bar{l},\bar{p}]$,
$[\bar{n},0,0,0] \to [\bar{n},0,1,0] \to$ F.

The boundary between $[\bar{n},0,0,0]$ and $[\bar{n},0,1,0]$ is
non-degenerate and cannot be split by terms of higher order in
$D^{-1}$. The boundary between  $[\bar{n},0,1,0]$ and F remains
degenerate to all phases of the form $[\bar{n},m,1,m]$. The effect of
higher order terms can be deduced by noting that $V_{\beta_1}(2m)$ and
$V_{\beta_2}(2p)$ are positive and convex. This implies that all the
transitions
$[\bar{n},m,1,m] \to [\bar{n},m+1,1,m+1]$ are stable [\onlinecite{FSzI}]. Fig.
5 summarises the results of the two-interface interaction analysis.

\section{Hyperfine Structure}

We now restrict our analysis to the richest region of the phase
diagram, i.\/e.\/ where $l=1$.
We already know that, in the two-interface approximation, the possible
ground states can be written in the form $\langle \alpha^n I_1 \beta_1^m I_2
\beta_2^m I_3 \rangle \equiv [n,m,m]$, where $I_1 \equiv 4$, $I_2
\equiv 030$, and $I_3 \equiv 2$.
Bassler, Sasaki and Griffiths [\onlinecite{Bass}] have shown that for
exponentially decaying interactions such as is the case here
the general form of the
interaction energy of an
arbitrary number of interfaces can be constructed as
\begin{eqnarray}
& &V_{\alpha \beta_1 \beta_2 \dots \beta_2}(2n,2m,2p,2q,\dots,2s)=
\nonumber \\
& & \ \ \ \ V_{\alpha}(2n) t_{\alpha\beta_1} V_{\beta_1}(2m) t_{\beta_1
\beta_2} V_{\beta_2}(2p)
t_{\beta_2 \alpha} V_{\alpha}(2q) \cdots t_{\beta_1 \beta_2} V_{\beta_2}(2s)
\label{eqn:factor}
\end{eqnarray}
\noindent where the $V$'s are defined in (\ref{eqn:V}) and
(\ref{eqn:Vbeta}) and to leading order
\begin{eqnarray}
t_{\beta_1 \beta_2} &=& D^{-3} \bigl\{ {\rm s}_2 -{\rm
s}_3 \bigr\}^{-2}c_2^2 c_3^2, \nonumber \\
t_{\alpha \beta_1} &=& D  \bigl\{ ({\rm s}_3 -{\rm
s}_2)({\rm s}_4 -{\rm s}_3 ) \bigr\}^{-1} ,\nonumber \\
 t_{\beta_2 \alpha} &=&t_{\alpha \beta_1}.
\label{eqn:t}
\end{eqnarray}

The formulae (\ref{eqn:t}) follow from calculations similar to that
described in Appendix A. For example taking the phases $A=[n,m,m]$,
$B=[n,m+1,m+1]$, and $C=[n,m,m,n,m+1,m+1]$ the right-hand side of
equation (\ref{eqn:short}) is equal in leading order to
$V_{\beta_1}(2m) t_{\beta_1\beta_2} V_{\beta_2}(2m+2)$.
$t_{\beta_1\beta_2}$ can then immediately be extracted by using the expression
(\ref{eqn:Vbeta}) for $V_{\beta_1}$ and $V_{\beta_2}$.

With the aid of (\ref{eqn:factor}) it is
possible to examine the effects of three-interface interactions on the
superdegenerate boundaries in Fig. 5. Consider the general case represented in
Fig.\ \ref{fig:cross}(a).
All four boundaries are multiphase lines where any sequence of the two
neighbouring phases are degenerate within the two-interface interaction
approximation. For the $[n,m,m]:[n+1,m,m]$  and $[n,m+1,m+1]:[n+1,m+1,m+1]$
boundaries this exhausts the possibilities and the three-interface interactions
are not of sufficiently long range to split the degeneracy.
For the $[n,m,m]:[n,m+1,m+1]$
($[n+1,m,m]:[n+1,m+1,m+1]$) boundary, however, the phases $[n,m+1,m]$ and
$[n,m,m+1]$  ($ [n+1,m+1,m]$ and $[n+1,m,m+1] $) are
also degenerate and there is the possibility that these may be
stabilised with respect to $[n,m,m]$ and $[n,m+1,m+1]$
( $[n+1,m,m]$ and $[n+1,m+1,m+1]$)
by the three-interface interaction.

To check this we need
the energy differences

\begin{eqnarray}
\lefteqn{2 E_{[n,m+1,m]} - E_{[n,m,m]}- E_{[n,m+1,m+1]} = 2 E_{[n,m,m+1]} -
E_{[n,m,m]}- E_{[n,m+1,m+1]}  } \nonumber \\
& &\sim V_{\beta_1 \beta_2} (m, m+1) + V_{\beta_1 \beta_2} (m+1, m)
-V_{\beta_1 \beta_2} (m+1, m+1) - V_{\beta_1 \beta_2} (m, m)
\end{eqnarray}
which are dominated by $V_{\beta_1 \beta_2} (m, m)$ and which are
therefore negative. Similarly the $[n+1,m,m]:[n+1,m+1,m+1]$ boundary
is unstable with
respect to the formation of $\{ [n+1,m+1,m], [n+1,m,m+1] \}$.
The resulting modification to the phase diagram is
sketched in Fig.\ \ref{fig:cross}(b).

The $V_4$  terms do not cause further splitting of the multidegenerate
lines of Fig.\ \ref{fig:cross}(b) but they qualitatively change the
phase diagram near the two points where four lines meet.
In proximity of the upper one the structure of the phase diagram is
determined by the signs of the energy differences [\onlinecite{Bass}]
\begin{eqnarray}
\Delta V_1 &=& V_{\alpha \beta_1 \beta_2}(n,m,m)
+V_{\alpha \beta_1 \beta_2}(n+1,m+1,m)
-V_{\alpha \beta_1 \beta_2}(n+1,m,m)
-V_{\alpha \beta_1 \beta_2}(n,m+1,m)
\nonumber \\
\Delta V_2 &=& V_{\beta_1 \beta_2 \alpha}(m,m,n)
+V_{\beta_1 \beta_2 \alpha }(m+1,m,n+1)
-V_{\beta_1 \beta_2 \alpha }(m+1,m,n)
-V_{\beta_1 \beta_2 \alpha }(m,m,n+1) \nonumber \\
{}~&~&~
\label{eqn:differences}
\end{eqnarray}

With the aid of the factorization formul\ae\/ (\ref{eqn:factor}) it is
possible to check that the two energy differences (\ref{eqn:differences}) are
positive. This means that phases $[n+1,m,m]$ and
$\{ [n,m,m+1],[n,m+1,m] \}$ are separated by a short
first-order line; similarly one can show that $\{
[n+1,m,m+1],[n+1,m+1,m] \}$ and $[n,m+1,m+1]$ also coexist at
a first-order transition. In this approximation the structure of
Fig. \ref{fig:cross}(b) must be modified as in Fig. \ref{fig:cross}(c).

The factorization formul\ae\/ (\ref{eqn:factor}) allow
us to go further and study the the effect on the phase diagram
of interface--interface interactions of
all orders. Bassler, Sasaki and Griffiths [\onlinecite{Bass}]
showed that the form of the phase diagram depends upon the sign of the
two-interface interactions (\ref{eqn:V}) and (\ref{eqn:Vbeta}) and the $t$'s,
equation (\ref{eqn:t}).
Here these are all positive corresponding to a case where
the superdegenerate
boundaries at the end of the first-order lines in Fig. 6(c) split
under the effect of higher-order interface--interface interactions,
giving rise to a structure analogous to that in Fig. 5 (but where
the phases have longer periodicity). Furthermore one can carry the
analysis further by studying again the splitting near the points where four
lines meet and so on, finding a structure similar to the one in Fig. 6(c).
The analysis can then be repeated ad infinitum, showing that the
$\Upsilon$-point has, indeed, a self-similar, fractal structure.

%

\section{Discussion}

The analysis presented above was based on retaining only the
leading order term in the interface--interface interaction.
We cannot rule out the possibility that the
neglected higher-order contributions
could affect the phase diagram.
In particular there will be correction terms ${\cal O}(l^2/D^2)$ where
$l$ is the period of a given phase which could introduce qualitative
changes for $l$ large and $D$ not sufficiently small.

The results were checked numerically in two ways. Firstly we used the
Floria--Griffiths algorithm [\onlinecite{FG}]
on a grid of size 1200
to check which phases
appeared. It was possible to resolve phases with $l$ up to 13. Secondly we
used a mean-field analysis, exact at zero temperature, to check the
positions of the phase boundaries. In this way the the formul\ae\ for the
interface--interface interactions could be verified for
short-period phases (typically $l$ up to 9).

To summarise, we have presented analytic evidence
for the existence of an $\Upsilon$-point
in a spin model. The phase diagram has been constructed inductively by
calculating the interface--interface interactions to leading order in
$1/D$, the inverse spin anisotropy.
Following arguments due to Bassler, Sasaki and Griffiths [\onlinecite{Bass}]
we have argued that
the $\Upsilon$-point has a self-similar, fractal structure.

\acknowledgements
JMY acknowledges the support of an EPSRC Advanced Fellowship and CM an
EPSRC Studentship and a grant from the Fondazione "A. della Riccia", Firenze.
We thank Prof.\ R.B.\ Griffiths for helpful discussions.

\appendix

\section{Calculation of the two-interface interaction.}

As an example of how to obtain the two-interface interaction we
consider explicitly the calculation of $\tilde{V}_{\alpha}(6)$.
Following Fig. 4 we need to consider the periodic phases
listed below where $n_1=4$, $N=24$ and $n=n_I=3$.
 A choice of labelling that satisfies (\ref{eqn:JDelta1}) for
$1 \le i \le n_1$ and
(\ref{eqn:JDelta2}) for $n_1+1 \le i \le N$ is as shown.

\begin{small}
\begin{center}
\begin{tabular}{ l l l l l l l l l l l l l l l l l l l l l l l l l l l l l
l l l l}
& & & & &\  &$i$ &= &$n_1;$ &1, &2 &...&  &    & & & &  & \\
& & &  & \  &\ & &   &  $\downarrow$ &$\downarrow$   & \\
A:\ &\ \  & & \multicolumn{3}{l}{${3 {\rm a}^0_i / \pi}$}  \ &\ &\ &  5 & 1 & 5
&1 &\\ \\
& & &  & \  &\ &$i$ &= & &...&N;&\multicolumn{3}{l}{$n_1+1$}&... & &  & &  &  &
 & & & & & & & \\
& & & & \  &\ & & & & &$\downarrow$& $\downarrow$  & &  & &  &  &
 & & & & & & & & \\
B:\ &\ &\ & \multicolumn{3}{l}{${3 {\rm b}^0_i / \pi}$}  \ &\ &\ & 5 & 1 &5 &1
& 5 & 1 & 4 & 0 & 2&
5 & 1 & 5 & 1 & 5 & 1 &5 & 1 & 4 & 0 & 2 & \\ \\

& & & &\ &\ &$i$ &= &N; &1  &...&  &$n_1$,  & \multicolumn{3}{l}{$n_1+1$}
&... & & &  & \\
& & & &\ &\  & &   &  $\downarrow$ &$\downarrow$   & &  &$\downarrow$
&$\downarrow$
& &  \\
C:\ &\ &\ &\multicolumn{3}{l}{$3 {\rm c}^0_i/ \pi$} & \  &\ &  5 & 1 & 5 &1 & 5
& 1 &5 &1 & 5 & 1 & 4 & 0 & 2&
5 & 1 & 5 & 1 & 5 & 1 &5 & 1 & 4 & 0 & 2  \\
\end{tabular}
\end{center}
\end{small}
\begin{equation}
{}~~ \label{eqn:table}
\end{equation}

We can now use (\ref{eqn:short}) to calculate
$\tilde{V}_{\alpha}(6)$ to leading order. The quantities $(\tilde{a}_{n_1} -
\tilde{b}_N)$, $(\tilde{a}_1 - \tilde{b}_{n_1+1})$, $(\tilde{c}_1 -
\tilde{c}_{n_1+1})$ and  $(\tilde{c}_N - \tilde{c}_{n_1})$
can be obtained correct to leading order from the linear
approximation to the recursion equation (\ref{eqn:rec})

\begin{equation}
\tilde{\theta}_i= \biggl\{
 - 2 h_i^\theta
(\epsilon_i^\theta + \tilde{\theta}_i)
+2 J_{i-1,i}^\theta ( \tilde{\theta}_{i-1} - \tilde{\theta}_i
+ \Delta_{i-1,i}^\theta)
-2 J_{i,i+1}^\theta ( \tilde{\theta}_{i} -\tilde{\theta}_{i+1}
 +\Delta_{i,i+1}^\theta) \biggr\}/D
\label{eqn:recexpl}
\end{equation}
where we have used the definitions
(\ref{eqn:first})--(\ref{eqn:last}).

Let
\begin{equation}
\tilde{\theta}_i = { \theta^1_i \over D} +  { \theta^2_i \over D^2} +
...\ .
\end{equation}
\noindent Substituting into (\ref{eqn:recexpl}) and equating like powers
of $D^{-1}$ gives
\begin{eqnarray}
\theta^1_i &=&  2 J_{i-1,i}^\theta \Delta_{i-1,i}^\theta
-2 J_{i,i+1}^\theta \Delta_{i,i+1}^\theta - 2 h_i^\theta
\epsilon_i^\theta, \label{eqn:recu1} \\
\theta^n_i & = & - 2 h_i^\theta  {\theta}_i^{n-1}
-2 J_{i-1,i}^\theta ( {\theta}_{i-1}^{n-1} - {\theta}_i^{n-1})
-2 J_{i,i+1}^\theta ( {\theta}_{i}^{n-1} -{\theta}_{i+1}^{n-1}), \ \ \ \ \
n>1. \label{eqn:recu2}
\end{eqnarray}

To calculate $(\tilde{a}_1 - \tilde{b}_{n_1+1})$ it is helpful to
display explicitly $a_i^0$ and $b_{n_1+i}^0$ as a function of $i$.

\begin{small}
\begin{center}
\begin{tabular}{ l l l l l l l l l l l l l l l l l l l l l l l l l l l l}
 &  &$i$ & & &... & & -3 & -2 & -1  & 0 & &1&2& 3 &4 &... & & &  &  &
& & & & &\\
  & \ &\ &\ & & & &* & & & &  &  &  &  & * & &
 &  &  &  &  &  & &  &\\
 \multicolumn{3}{l}{$3 \rm{a}^0_i / \pi$} & \ &\  &... &5&1&5&1 & 5 &:& 1 & 5 &
1 & 5 &1&5&1 &...\\
 \multicolumn{3}{l}{$3 b^0_{n_1+i}/ \pi $} & \ &\  &... &0 & 2 & 5 & 1
& 5 &:& 1 & 5 & 1 & 4 & 0 & 2&
5 & ... \\
 & \ &\ &\ & & & &* & & & & &  &  &  & * & &
 &  &  &  &  &  & &  &\\
\end{tabular}
\end{center}
\end{small}
\begin{equation}
{}~~\label{eqn:nono}
\end{equation}

The $*$'s mark where $a^0_i$ first differs from
$b^0_{i+n_1} $ when moving away from the dotted interface in either
direction.

It follows immediately from (\ref{eqn:recu1}) that
\begin{equation}
(\tilde{a}_3-\tilde{b}_{n_1+3})= {1 \over D} \biggl\{ \sin[ {\pi \over
3} \Delta -3] - \sin[ {\pi \over 3} \Delta -4] \biggr\}
\end{equation}
\noindent Two further iterations of (\ref{eqn:recu2}) give
\begin{eqnarray}
(\tilde{a}_1-\tilde{b}_{n_1+1})&=&{1 \over D^2} \cos [ {\pi \over 3}
\Delta + a^0_1 - a^0_2] \cos [ {\pi \over 3} \Delta + a^0_2 - a^0_3]
(\tilde{a}_3-\tilde{b}_{n_1+3})\nonumber \\
 &=&{1 \over D^3} \cos [ {\pi \over 3}
(\Delta-4) ] \cos [ {\pi \over 3} (\Delta -2)] \biggl\{ \sin[ {\pi \over
3} \Delta -3] - \sin[ {\pi \over 3} \Delta -4] \biggr\}.
\label{eqn:a1}
\end{eqnarray}

\noindent $(\tilde{a}_{n_1}-\tilde{b}_{N})$ may be calculated in an
analogous way
\begin{equation}
(\tilde{a}_{n_1}-\tilde{b}_{N}) =-(\tilde{a}_1-\tilde{b}_{n_1+1}).
\label{eqn:a2}
\end{equation}
Similarly
\begin{equation}
(\tilde{c}_1-\tilde{c}_{n_1+1}) = (\tilde{c}_N-\tilde{c}_{n_1})
= (\tilde{a}_1-\tilde{b}_{n_1+1}).
\label{eqn:bs2}
\end{equation}

\noindent Using $J_{0,1}=\cos [ {\pi \over 3} (\Delta -2)]/2$, from the
definition (\ref{eqn:first}) and
substituting (\ref{eqn:a1})--(\ref{eqn:bs2}) into (\ref{eqn:short})  gives
\begin{eqnarray}
\tilde{V}_{\alpha}(6) &=& -{1 \over D^6} \biggl\{ \sin[ {\pi \over
3} \Delta -4] - \sin[ {\pi \over 3} \Delta -3] \biggr\}^2 \cos [ {\pi
\over 3} (\Delta -4)]^2
 \cos [ {\pi \over 3} (\Delta -2)]^3.
\label{eqn:final}
\end{eqnarray}

It is important to point out that the labelling used in equation
(\ref{eqn:table}) is not
unique. Any labelling which satisfies (\ref{eqn:JDelta1}) for $1 \le i
\le n_1$ and (\ref{eqn:JDelta2}) for $1 \le n_1+1 \le N$ will give the
correct results for $\tilde{V}_\alpha(6)$. However in general
$(\tilde{a}_1 - \tilde{a}_{n_1+1})$ etc. will contain lower powers of
$1/D$ which cancel when the difference in  (\ref{eqn:short}) is
taken. The choice
given above, which maximises the distance of the position (*) where
${\rm a}^0_i$ first differs from ${\rm b}^0_{n_1+1}$ avoids such
cancellation and leads to the easiest calculation.

Finally we mention that because the
interface-interface interactions decay very rapidly
(exponentially) with increasing interface--interface distance
the values of $n_1+2n$ and \linebreak
$N-2n-2n_I-n_1$ need not in fact be much larger that $2n$ and $n_I$.
That sufficiently large values have been chosen can be checked {\em a
posteriori} by verifying that increasing the values of $n_1$ and $N$ does
not change the result (\ref{eqn:final}).

\newpage

\begin{figure}
\centerline{\psfig{figure=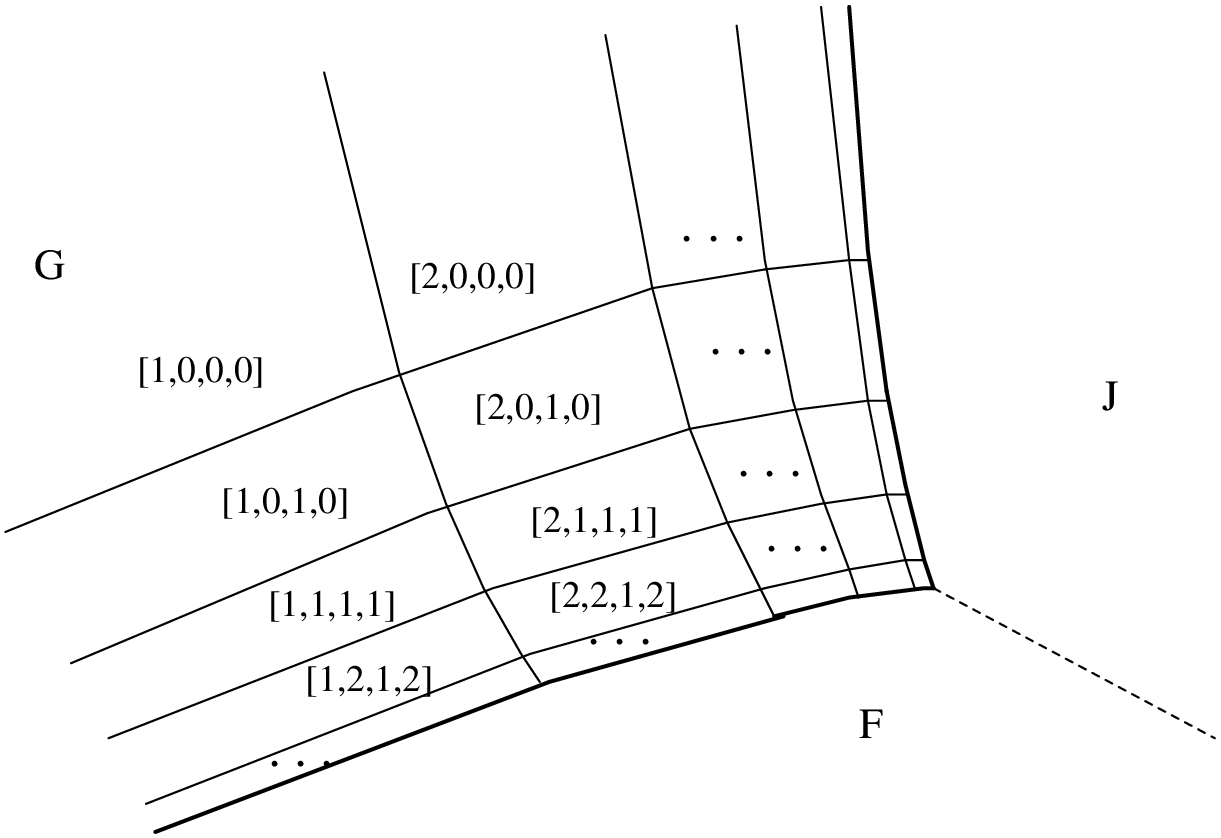,width=3.5in,height=2.0in}}
\vspace{0.3cm}
\caption{Schematic phase diagram near the $\Upsilon$-point in the two-interface
interaction
approximation. The notation $[n,m,l,p]$ is used to indicate the phase
$\langle (51)^n 4 (14)^m 0 (30)^l 2 (52)^p \rangle$. The bold solid
lines are accumulation lines. The dashed line is a first order boundary.}
\label{fig:upsi2}
\end{figure}

\begin{figure}
\centerline{\psfig{figure=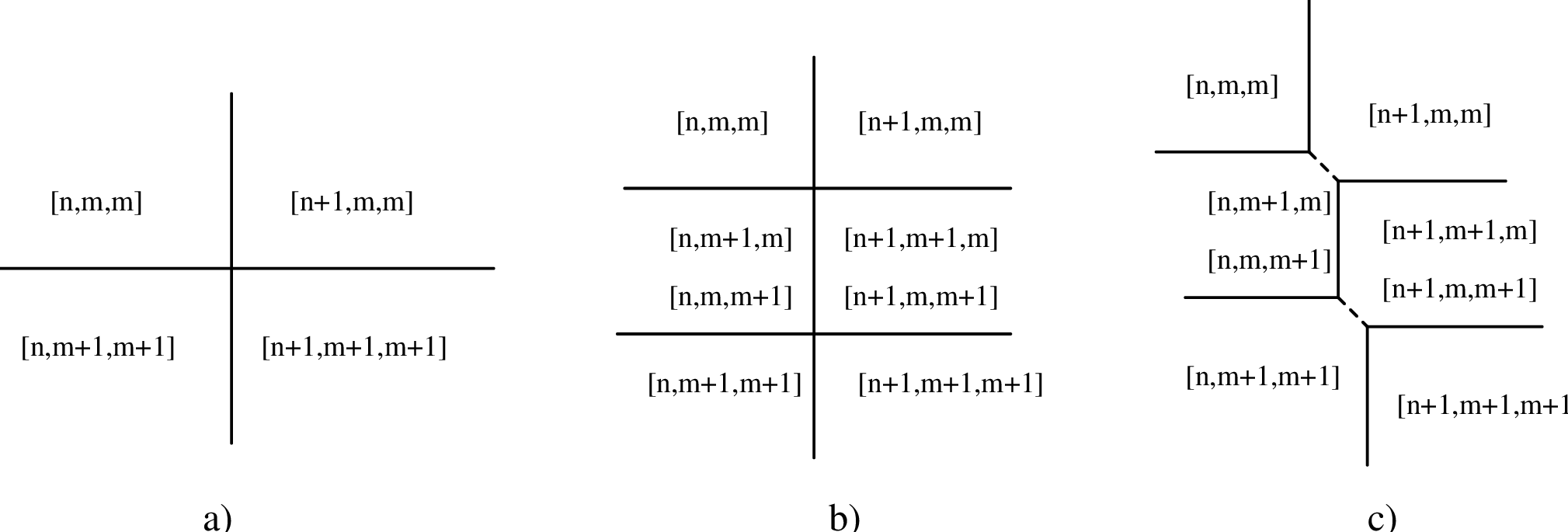,height=2.0in }}
\vspace{0.3cm}
\caption{Detail of the phase diagram in (a) the two-interface
approximation; (b) the three-interface approximation; (c) the
four-interface approximation. First order lines are dashed.
}
\label{fig:cross}
\end{figure}

\end{document}